\documentclass[pre,aps,amsmath,amssymb,reprint,floatfix,superscriptaddress]{revtex4-1}

\pdfoutput=1
\usepackage{xcolor}
\usepackage{siunitx}
\usepackage{graphicx}
\usepackage[colorlinks=true,citecolor=blue,urlcolor=blue]{hyperref}
\usepackage{etoolbox} 
\usepackage{bm} 
\usepackage[normalem]{ulem}

\def\eqq#1{Eq.~(\ref{#1})}

\def\eq#1{(\ref{#1})}

\def\f#1{Fig.~\ref{#1}}

\def\s#1{Section~\ref{#1}}
\def\a#1{Appendix~\ref{#1}}

\def\c#1{~\cite{#1}}

\def\cc#1{~Ref.~\cite{#1}}
\def\ccc#1{~Refs.~\cite{#1}}

\def\av#1{\langle #1 \rangle}

\def\beq{\begin{equation}}
\def\eeq{\end{equation}}
\def\bea{\begin{eqnarray}}
\def\eea{\end{eqnarray}}

\def\e{{\rm e}}
\def\x{{\bm x}}
\def\cee{{\bm c}}

\def\kB{k_{\rm B}}
\def\tf{t_{\rm f}}
\def\kt{\kB T}

\definecolor{blue}{rgb}{0,0,0}
\newcommand{\bb}[1]{\textcolor{blue}{#1}}

\begin{document}

\title{Improving noisy free-energy measurements by adding more noise}

\author{Stephen Whitelam}
\email{swhitelam@lbl.gov}
\affiliation{Molecular Foundry, Lawrence Berkeley National Laboratory, 1 Cyclotron Road, Berkeley, CA 94720, USA}

\begin{abstract}
Estimating free-energy differences using nonequilibrium work relations, such as the Jarzynski equality, is hindered by poor convergence when work fluctuations are large. For systems governed by overdamped Langevin dynamics, we propose the counterintuitive approach of {\em adding} noise in order to increase the precision of such calculations. By introducing additional stochastic fluctuations to the system and rescaling its potential energy \bb{accordingly}, we leave the thermodynamics of the system unchanged while increasing its relaxation rate. For a given time-dependent protocol this modification reduces \bb{the} dissipated \bb{reduced} work, leading to more accurate free-energy estimates. \bb{The method is designed to be used in experiment, and we illustrate its operation} using computer simulations applied to two model systems. However, the regime of applicability of this strategy is likely limited, because it requires control of the system's potential energy in a way that is feasible in only a few experimental settings.
\end{abstract}

\maketitle

\section{Introduction}

The Jarzynski equality and the Crooks fluctuation theorem provide a way to estimate equilibrium free-energy differences from nonequilibrium work measurements \cite{jarzynski1997nonequilibrium,crooks1999entropy}. These relations apply to systems initially at equilibrium and driven out of it by the variation of a set of external control parameters. However, such estimates can be numerically challenging. The Jarzynski equality involves an exponential average, and rare trajectories with atypically small work values contribute disproportionately, making free-energy estimates unreliable unless a large number of trajectories are sampled \cite{jarzynski2006rare,hummer2010free,rohwer2015convergence}. This problem worsens as the rate of driving increases, because the mean dissipated \bb{reduced} work $\beta(\langle W \rangle - \Delta F)$ grows~\footnote{The adjective {\em reduced} refers to the multiplication of a quantity by $\beta$.}, leading to an exponential increase in the number of trajectories required to estimate the free-energy difference to a given precision. \bb{Here $\beta \equiv 1/(\kt)$.} Various strategies have been explored to mitigate this issue, including optimizing control protocols to minimize dissipated work and work fluctuations\c{schmiedl2007optimal,geiger2010optimum,vaikuntanathan2008escorted,blaber2020skewed,arrar2019accurate,loos2024universal,zhong2022limited}. 

In this paper we propose an alternative and counterintuitive approach to increasing the precision of free-energy measurements for systems governed by overdamped Langevin dynamics: add noise. By injecting additional stochastic fluctuations into the system, and rescaling its potential energy \bb{accordingly}, we can increase the system's relaxation rate without altering its thermodynamics. For fixed driving rate, this modification reduces the \bb{reduced dissipated} work and so suppresses rare-event sampling problems, leading to more precise estimates of $\Delta F$. We demonstrate this effect in two model systems, using numerical simulations, \bb{and show} that this noise-engineering strategy can significantly enhance the accuracy of free-energy calculations. Our results \bb{indicate} that noise engineering could be a valuable tool for enhancing thermodynamic measurements and computations, and \bb{suggest} that the notion of optimal control\c{alvarado2025optimal} could be broadened to include control of noise, which is now experimentally feasible\c{martinez2012effective,chupeau2018thermal,saha2023information,aifer2024thermodynamic}, as well as control of the potential. 

\bb{Our intent is that this method be applied to experiments, and we use numerical simulations to illustrate its operation (see\ccc{vaikuntanathan2008escorted,sivak2013using,wu2020stochastic,zhong2024time} for approaches designed to enhance sampling and precision in numerical settings)}. However, the regime of applicability of this method to experimental free-energy calculations is limited. Noise-injection methods have been demonstrated in various experimental settings, but the method discussed here also depends on our ability to control the system's potential energy, which is feasible in only a few experimental settings. We describe the method in \s{tools}, and apply it to two model systems in \s{results}. We conclude in \s{conclusions}, where we comment on the applicability of the approach to various experimental systems.

\section{Free energies from nonequilibrium measurements} 
\label{tools}

Consider a system described by $N$ microscopic coordinates $\x =\{x_i\}, \, i=1,2,\dots ,N$, possessing an energy function $U(\x,\cee(t))$. The vector $\cee(t)$ specifies the time evolution of a set of control parameters that will be used to drive the system out of equilibrium. The system is in contact with a heat bath at temperature $T$ (with $\beta \equiv 1/(\kt)$, where $k_{\rm B}$ is Boltzmann's constant), and is initially in thermodynamic equilibrium with respect to the energy function $U(\x,\cee(0))$. 

The system's microscopic coordinates evolve according to the overdamped Langevin equation\c{van1992stochastic,ciliberto2017experiments}
\beq
\label{lang}
\dot{x}_i = -\mu \frac{\partial}{\partial x_i} U(\x,\cee(t)) + \sqrt{2 \kt \mu} \, \eta_i(t),
\eeq
on the time interval $t \in [0,\tf]$. Here $\mu$ is the mobility parameter, which sets the basic time scale of the system, and $\eta_i$ is a Gaussian white noise with mean $\av{\eta_i(t)}=0$ and covariance $\av{\eta_i(t) \eta_j(t')}= \delta_{ij} \delta(t-t')$. The noise $\eta$ represents the thermal fluctuations inherent to the system.

The dynamics \eq{lang} satisfies detailed balance with respect to the temperature parameter $\beta$ and the energy function $U$, and so if the system is driven out of equilibrium by the application of a time-dependent protocol $\cee(t)$ then it obeys the Crooks\c{crooks1999entropy} and Jarzynski\c{jarzynski1997nonequilibrium,jarzynskia2008nonequilibrium} relations
\beq
\label{fluc}
P_{\rm F}(W) \hspace{0.3pt} \e^{-\beta W}=  \e^{-\beta \Delta F} P_{\rm R}(-W),
\eeq
and
\beq
\label{jarz}
\av{\e^{-\beta W}}=\e^{-\beta \Delta F}.
\eeq
In these equations $W$ is the work done to enact the protocol $\cee(t)$,
\beq
W=\int_0^{\tf} {\rm d} t \, \frac{{\rm d} \cee(t)}{{\rm d} t} \cdot \left(\frac{\partial U(\x, \cee(t))}{\partial \cee}\right)_{\bm x} ;
\eeq
$\Delta F$ is the free-energy difference associated with the initial and final values of the energy function, namely
\beq
\label{df}
\Delta F = -\kt \ln \frac{\int {\rm d} \x \, \e^{-\beta U(\x,\cee(\tf))}}{\int {\rm d} \x \, \e^{-\beta U(\x,\cee(0))}};
\eeq
 $P_{\rm F}(W)$ denotes the probability distribution of work under the forward protocol $\cee(t)$; $P_{\rm R}(-W)$ denotes the probability distribution of the negative of the work under the time-reversed protocol $\cee(\tf-t)$ (starting in equilibrium with respect to the potential $U(\x, \cee(\tf))$); and the average $\av{\cdot}$ is taken over many independent realizations of the forward protocol. 
 
Equations \eq{fluc} and \eq{jarz} allow the extraction of the equilibrium free-energy difference $\Delta F$ from the work measured during a set of nonequilibrium experiments. For instance, we can calculate the Jarzynski estimator for $\beta \Delta F$, 
\beq
\label{jarz1}
J = -\ln \left(N_{\rm traj}^{-1} \sum_{i=1}^{N_{\rm traj}} \e^{-\beta W_i}\right),
\eeq
where $i$ labels trajectories, as well as a measure of the statistical error in this quantity, the variance $\sigma^2_{B}$ of the \bb{block average $B$ of the Jarzynski estimator. The $j^{\rm th}$ block average
\beq
\label{block}
B_j =N_{\rm block}^{-1} \sum_{i=1}^{N_{\rm block}} e^{-\beta W_{j,i}}
\eeq
is computed from $N_{\rm block} = 100$ trajectories, where $W_{j,i}$ is the work associated with trajectory $i$ in block $j$. We estimate the variance $\sigma_{\rm B}^2$ of the block average as
\beq
\sigma_B^2 = N_{\rm samples}^{-1} \sum_{j=1}^{N_{\rm samples}} B_j^2 - \overline{B}^2,
\eeq
where 
\beq
\overline{B} = N_{\rm samples}^{-1} \sum_{j=1}^{N_{\rm samples}} B_j, 
\eeq
and $N_{\rm samples} = 10^4$. The virtue of the block average is that it provides a straightforward way to estimate the statistical uncertainty of the Jarzynski estimator without requiring assumptions about the underlying distribution of $W$. By aggregating trajectories into independent blocks, we obtain a distribution of values $B$ from which the variance can be directly calculated.}

\bb{We can also calculate the number of trajectories $N(\Delta f)$ required to estimate $\Delta F$ to a precision of $\Delta f \, k_{\rm B}T$, using the formula\c{geiger2010optimum}
\beq
 \label{number}
N_{\Delta f} = \frac{1}{(\Delta f)^2} \left( \left\langle \e^{-2 \beta W} \right\rangle - 1 \right).
\eeq}

In \s{results} we will show for two model systems that rapidly-driven protocols allow only a rough estimate of $\Delta F$.

One way to increase the precision with which we estimate $\Delta F$ is to increase the trajectory length $\tf$ and carry out the protocol more slowly. Doing so reduces the dissipated work, which in general allows better convergence of \eq{jarz} (the illustrative case of Gaussian work fluctuations is discussed in \a{gauss}).  An alternative and counterintuitive way to increase the precision with which we estimate the free-energy difference is to increase the noise present in the system, as we now describe.

If we rescale the energy function $U(\x,\cee(t))$ by a factor of $\lambda>1$, and inject into the system Gaussian white noise with variance $\sigma^2 = 2 \kt \mu (\lambda-1)$, the equation of motion \eq{lang} becomes 
\beq
\label{lang2}
\dot{x}_i = -\mu \frac{\partial}{\partial x_i} \lambda U(\x,\cee(t)) + \sqrt{2 \kt \mu} \, \eta(t)+\sigma \zeta_i(t),
\eeq
where $\av{\zeta_i(t)}=0$ and $\av{\zeta_i(t) \zeta_j(t')}=\delta_{ij} \delta(t-t')$.  The two noise terms in \eq{lang3} can be considered an effective Gaussian white noise of variance $2 \kt \mu \lambda$, and so \eq{lang2} can also be written
\beq
\label{lang3}
\dot{x}_i = -\mu \frac{\partial}{\partial x_i} U_\lambda(\x,\cee(t)) + \sqrt{2 k_{\rm B} T_\lambda \mu} \, \eta_i(t),
\eeq
where the correlations of $\eta$ are as previously. \eqq{lang3} is a modification of \eqq{lang}, and contains a scaled energy function $U_\lambda \equiv \lambda U$ and an effective temperature $T_\lambda \equiv \lambda T $ (and correspondingly an effective temperature parameter $\beta_\lambda \equiv \beta/\lambda$). 
\begin{figure*} 
   \centering
   \includegraphics[width=0.85\linewidth]{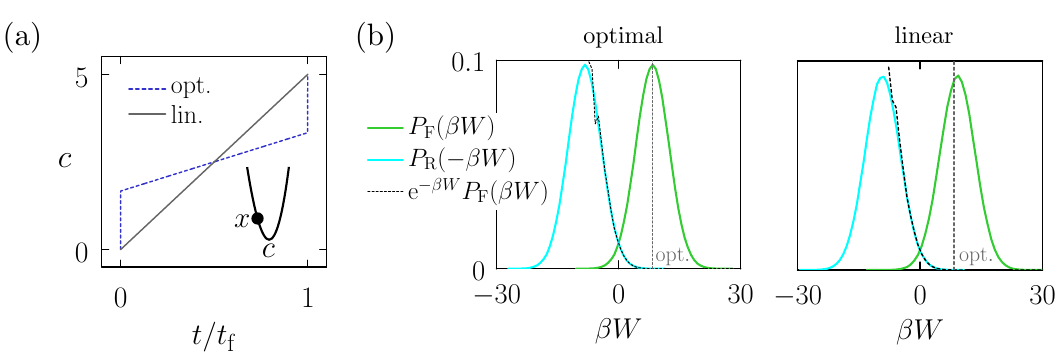} 
   \caption{(a) Work-minimizing (dotted) and linear (solid) protocols for the trap-translation problem of\cc{schmiedl2007optimal}. The trajectory length is $\tf=1$. Inset: sketch of the potential \eq{vee}. (b) Work distributions for these protocols and their time reverse, which obey the Crooks relation \eq{fluc}. The free-energy difference $\Delta F=0$ can be estimated from the crossing of the distributions. Distributions were calculated using $10^6$ independent trajectories. The vertical grey dashed line (opt) indicates the value of $\av{W}$ for the optimal protocol.}
   \label{fig1}
\end{figure*}

The work done carrying out the protocol $\cee(t)$ under the new dynamics is
\beq
W_\lambda=\int_0^{\tf} {\rm d} t \, \frac{{\rm d} \cee(t)}{{\rm d} t} \cdot \left(\frac{\partial U_\lambda(\x, \cee(t))}{\partial \cee}\right)_{\bm x}.
\eeq
The Jarzynski equality now reads 
\beq
\label{jmod}
\av{\e^{-\beta_\lambda W_\lambda}}_\lambda=\e^{-\beta \Delta F},
\eeq
where $\av{\cdot}_\lambda$ denotes an average over the $\lambda$-modified dynamics, \eqq{lang2}. (In previous equations, the average $\av{\cdot}$ corresponds to the original dynamics, i.e. to the choice $\lambda=1$.) 

\bb{Next, note that the exponent appearing in the Jarzynski relation is not the work $W$ but the {\em reduced} work $\beta W$. Accordingly, the} quantity in the exponential on the left-hand side of \eq{jmod} is $-\beta_\lambda W_\lambda$, appropriate for the scaled potential and effective temperature appearing in \eqq{lang3}. The resulting estimate of $\beta \Delta F$ is
\beq
\label{jarz2}
J_\lambda = -\ln  \left( N_{\rm traj}^{-1} \sum_{i=1}^{N_{\rm traj}} \e^{-\beta_\lambda (W_\lambda)_i} \right).
\eeq
Note that the right-hand sides of \eq{jarz} and \eq{jmod} are the same, because the dual modification of adding noise and scaling the potential leaves the thermodynamics of the system unchanged. Indeed, the combination $\beta_\lambda W_\lambda = \beta W$, and so $\beta_\lambda \av{W_\lambda}_\lambda= \beta \av{W}_\lambda$, although we will retain the $\lambda$ subscripts for clarity.

An alternative view of the modified dynamics can be obtained by rescaling time $t \to t/\lambda$ in \eqq{lang3}, giving
\beq
\label{lang4}
\dot{x}_i = -\mu \frac{\partial}{\partial x_i} U(\x,\cee(t/\lambda)) + \sqrt{2 \kt \mu} \, \eta(t),
\eeq
evolved on the time interval $ t \in [0, \lambda t_{\rm f}]$. \eqq{lang4} describes a system that involves the original energy function $U$ and temperature $T$ -- and so has the original thermodynamics -- but where the protocol $\cee$ is applied at a rate $\lambda$ times slower than previously, for a duration $\lambda$ times longer. 

This effective time-rescaled description suggests that the $\lambda$-modified dynamics will allow more accurate estimation of $\beta \Delta F$ using the Jarzynski identity than does the original dynamics. From our understanding of how work scales with trajectory duration\c{schmiedl2007optimal,van2021geometrical} we expect the dissipated \bb{reduced} work $\beta_\lambda \av{W_\lambda}_\lambda-\beta \Delta F$ to decrease with $\lambda$, and for large $\lambda$ to scale as $1/\lambda$. Reducing the dissipated \bb{reduced} work tends to reduce the variance of the Jarzynski estimator, and so we expect the statistics of $\av{\e^{-\beta_\lambda W_\lambda}}_\lambda$ to become increasingly better behaved as $\lambda$ increases. 

\bb{For example, for the illustrative case of Gaussian work fluctuations, discussed in \s{gauss}, a linear decrease with $\lambda$ in the dissipated reduced work results in a linear decrease with $\lambda$ in the variance of the dissipated reduced work. The latter results in an exponential decrease with $\lambda$ in the variance of the Jarzynski estimator, and hence an exponential decrease with $\lambda$ in the number of independent trajectories required to resolve the free-energy difference to a given precision.} 

Note that \eq{lang4} is an effective description. The protocol we apply, given in \eqq{lang2}, is the original protocol, and the duration of the experiment is the same as in the original dynamics. Instead, we have modified the original system by rescaling the potential and adding noise. 

\section{Numerical simulations} 
\label{results}

\subsection{Particle in a trap}
\begin{figure*} 
   \centering
   \includegraphics[width=\linewidth]{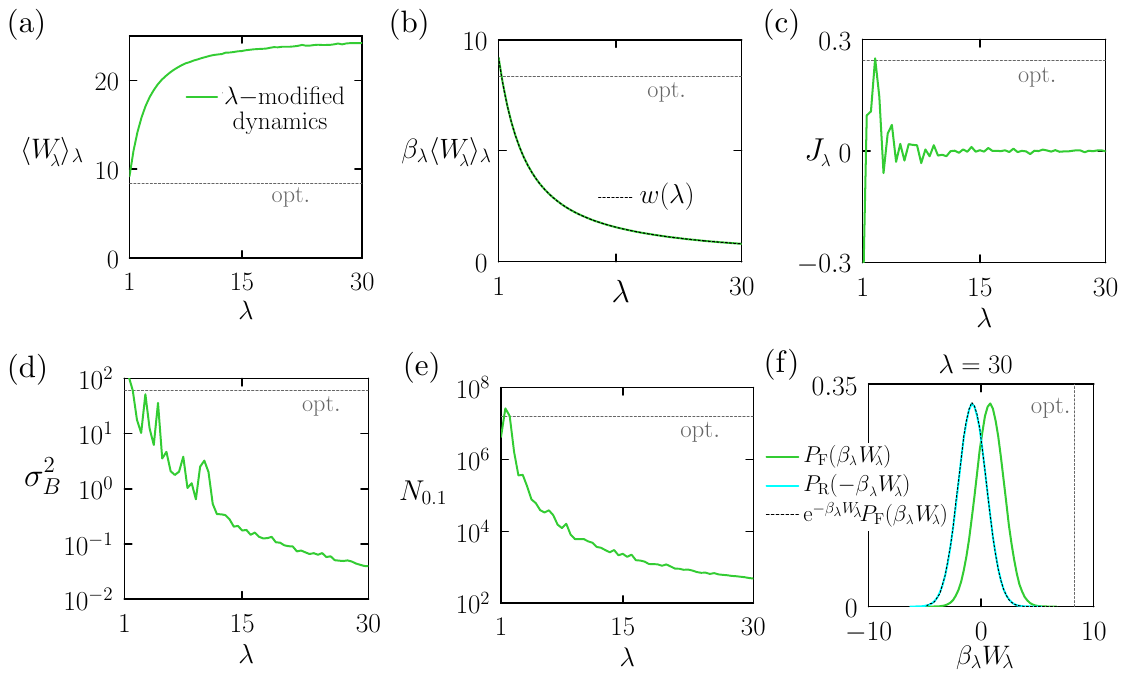} 
   \caption{Trap-translation model using the linear protocol within the $\lambda$-modified dynamics \eq{lang2} (green lines). The trajectory duration is $\tf=1$, as in \f{fig1}. Dashed lines labeled ``opt'' correspond to the optimal protocol used within the original dynamics (for which $\lambda=1$). (a) Mean work $\av{W_\lambda}_\lambda$, in units of $\kt$, as a function of $\lambda$. (a) Mean \bb{reduced} work $\beta_\lambda \av{W_\lambda}_\lambda$ as a function of $\lambda$. As expected from the arguments connecting \eqq{lang2} and \eqq{lang4}, this quantity behaves as $\beta \av{W}$ for the original dynamics, \eqq{lang}, if the protocol is imposed $\lambda$ times more slowly for $\lambda$ times longer. The function $w$ is given by \eqq{ww}. (c) Free-energy estimator \eq{jarz2}. (d) Variance of the block average \eq{block}, using $N_{\rm block}=100$ and $N_{\rm samples}=10^4$. (e) Number of trajectories \eq{number} required to estimate $\Delta F$ to a precision $0.1 \kt$; \bb{see \eqq{number}.} (f) Work distributions for forward and reverse protocols, for $\lambda=30$; compare the original dynamics of \f{fig1}(b). Averages and distributions are calculated using $N_{\rm traj} = 10^6$ trajectories.}
   \label{fig2}
\end{figure*}

\begin{figure} 
   \centering
   \includegraphics[width=\linewidth]{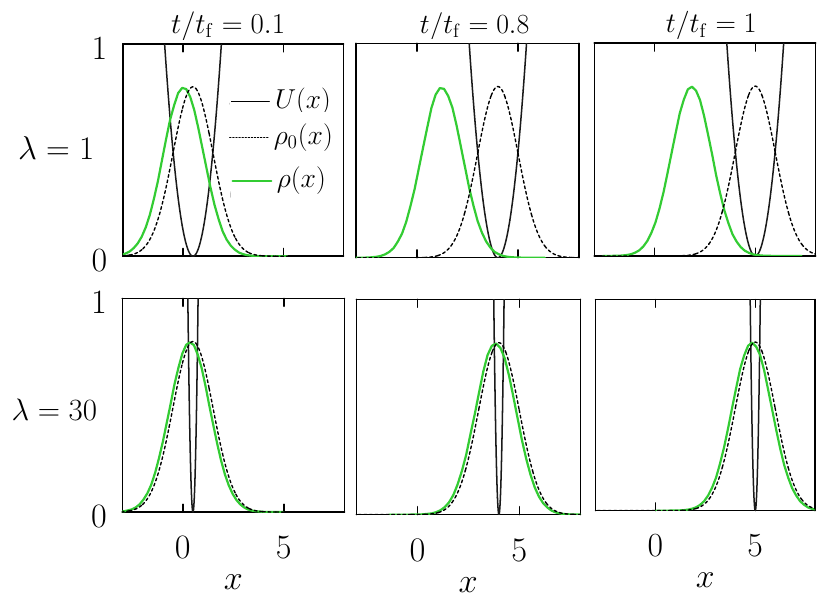} 
   \caption{Time-ordered snapshots for the trap-translation problem of\cc{schmiedl2007optimal}, using the linear protocol within the modified dynamics \eq{lang2}. The case $\lambda=1$ (top) corresponds to the original dynamics \eq{jarz}. The case $\lambda=30$ (bottom) corresponds to a system to which additional noise has been added and the trap potential rescaled. For the two cases we show the potential \eq{vee} (black), the associated Boltzmann distribution (black dashed), and the distribution of particle positions calculated using $10^6$ independent trajectories (green).}
   \label{fig3}
\end{figure}
 \begin{figure}[b] 
   \centering
   \includegraphics[width=\linewidth]{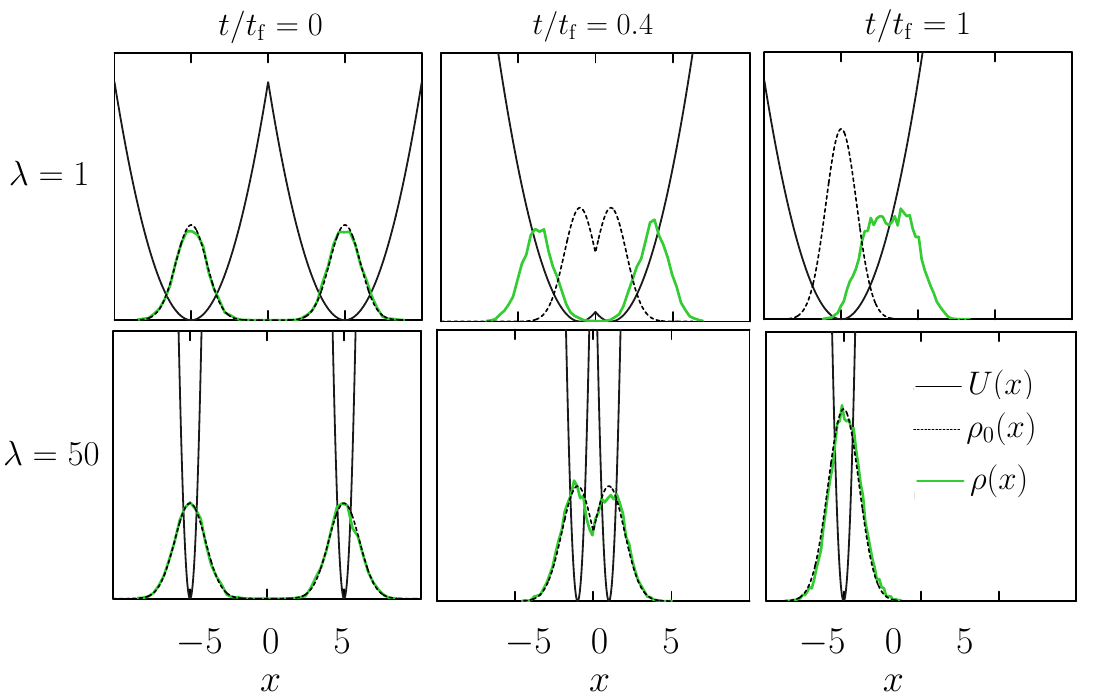} 
   \caption{Information erasure: a double-well potential is transformed to a single-well one, in time $\tf=1$, with a consequent increase in the free energy of the system of $\Delta F=\kt \ln 2$. We show time-ordered snapshots for the erasure protocol within the modified dynamics \eq{lang2}. The case $\lambda=1$ (top) corresponds to the original dynamics \eq{jarz}. The case $\lambda=50$ (bottom) corresponds to a system to which additional noise has been added and the potential scale increased. For the two cases we show the potential \eq{vee} (black), the associated Boltzmann distribution (black dashed), and the distribution of particle positions calculated using $10^5$ independent trajectories (green).}
   \label{fig_erase2}
\end{figure}

 \begin{figure*} 
   \centering
   \includegraphics[width=\linewidth]{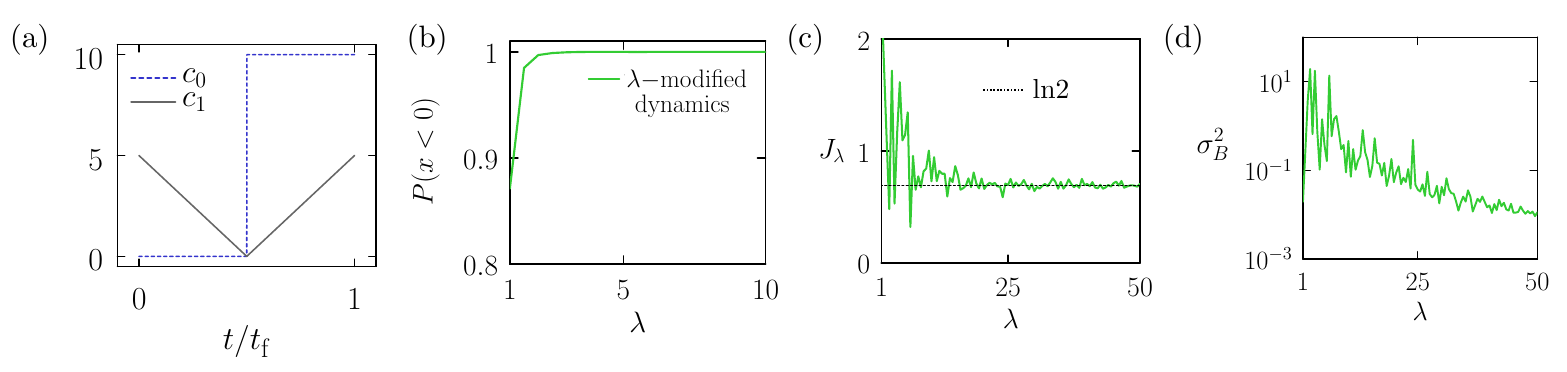} 
   \caption{Erasure using the basic protocol of\cc{Dago-2023-PNAS} within the $\lambda$-modified dynamics \eq{lang2}. (a) Time-dependent protocol $\cee(t)$. (b) Probability of successful erasure as a function of $\lambda$. (c) Free-energy estimator \eq{jarz2}. (d) Variance of the block average \eq{block}. Averages \bb{were} calculated using $10^5$ independent trajectories.}
   \label{fig_erase1}
\end{figure*}

To illustrate the numerical challenges of converging \eqq{jarz} when the timescale $\tf$ is small, consider the first problem of\cc{schmiedl2007optimal}, a model of a colloidal particle in a laser trap. A particle at position $x$ obeys the dynamics \eq{lang}, with the potential
\beq
\label{vee}
U(x,c(t))=\frac{k}{2} \left(x-c(t) \right)^2,
\eeq 
sketched in \f{fig1}. Here $k=1$, measured in units of $\kt$. The objective of the problem is to move the trap center $c(t)$ from an initial position $c(0)=0$ to a final position $c(\tf)=5$, in finite time $t_{\rm f}$, minimizing the work $\av{W}$ averaged over many realizations of the process. The free-energy change associated with the protocol is $\Delta F=0$. The protocol that minimizes mean work, which we will call the optimal protocol, has a linear form $c_{\rm opt}(t)=c_{\rm f} (t+1)/(t_{\rm f}+2)$ for $0<t<t_{\rm f}$, with jump discontinuities at the start $(t=0)$ and end $(t=t_{\rm f})$; see \f{fig1}(a). This protocol produces mean work \beq
\beta \av{W}_{\rm opt}=c_{\rm f}^2/(t_{\rm f}+2).
\eeq

\bb{For this system the work fluctuations are Gaussian, and so the work-minimizing protocol also minimizes work fluctuations and the error in estimating \eq{jarz}\c{geiger2010optimum}.}

We will also consider the linear protocol $c_{\rm lin}(t) = (t/\tf) c_{\rm f}$, shown as a solid line in \f{fig1}(a). This protocol produces mean work 
\beq
\label{ww}
\beta \av{W}_{\rm lin} = \frac{c_{\rm f}^2}{\tf^2} \left( \tf+\e^{-\tf} -1\right) \equiv w(\tf),
\eeq
which is slightly larger than that of the optimal protocol for all $\tf$\c{schmiedl2007optimal}. For large $\tf$, both $\beta \av{W}_{\rm opt}$ and $\beta \av{W}_{\rm lin}$ go to zero as $\sim c_{\rm f}^2/\tf$. 

Here we consider the case $\tf=1$, where the difference in mean work values between the two protocols is about 10\%, with $\beta \av{W}_{\rm lin} \approx 9.20$ and $\beta \av{W}_{\rm opt} \approx 8.33$.

\bb{The equation of motion for this system is linear stochastic differential equation with additive Gaussian white noise, making the trajectory $\{x(t)\}$ a Gaussian process. The total work performed during a protocol is a linear functional of this trajectory,
\beq
W[x(t)] = -k \int_0^{\tf} {\rm d}t\, (x(t) - c(t)) \dot{c}(t),
\eeq
and, because linear functionals of Gaussian processes are Gaussian random variables, the work distribution $P(W)$ is Gaussian, regardless of $c(t)$. For Gaussian work fluctuations, as shown in \s{gauss}, the variance of the work is related to its mean by the expression
 \beq
 \sigma^2=2 \kt \av{W}.
 \eeq
 For the linear protocol we therefore have $\sigma_{\rm lin}^2=2 \kt \av{W}_{\rm lin}$, with the mean given by \eqq{ww}.}
 
In \f{fig1}(b) we show the work distributions associated with the optimal and linear protocols (green), and with their time-reversed counterparts (cyan). Distributions were calculated using $N_{\rm traj}=10^6$ independent trajectories of \eqq{lang}, integrated using a first-order Euler scheme with timestep $10^{-5}$. The remaining parameters of the equation are $\kt=\mu=1$. Consistent with \eqq{fluc}, the distributions cross at a value of work $W$ approximately equal to $\Delta F=0$. We also show (black dashed) the forward distributions weighted by the quantity $\e^{-\beta W}$, which should be equal to the distributions associated with the time-reversed protocol. This is the case, but the comparison can be made over only about half the range of the reverse distribution.

  For the optimal protocol we measure $J= 0.24 \, \kt$, with $\sigma^2_{B} =59.709$ and $N_{0.1} =1.5 \times 10^7$~\footnote{The work distribution for the optimal protocol is Gaussian, and so $N_{0.1} = 100 ( \e^{2 \beta \av{W}} -1) \approx 1.7 \times 10^9$. Numerical calculation of the exponential average is highly imprecise for the pulling rate considered.}. For this choice of trajectory length, our estimate of $\Delta F$, using the optimal protocol, is accurate to only about a third of a $\kt$. For the linear protocol we measure $J= -0.45 \, \kt$, with $\sigma^2_{B} = 5652.6$ and $N_{0.1} = 0.4 \times 10^7$. 
 
We next show that the noise-injection procedure described in the previous section allows us to estimate $\Delta F$ for the trap system with better precision than can be achieved using the optimal protocol. In the $\lambda$-modified dynamics of \eqq{lang2} we use the linear protocol, because it is easier to quantify the notion of an effective time rescaling if the protocol is smooth. We use the same trajectory length ($\tf=1$) as before.

In \f{fig2} we show results obtained using the $\lambda$-modified dynamics (green lines), together with those obtained using the optimal protocol. Panel (a) shows that the mean work $\av{W_\lambda}_\lambda$ increases with $\lambda$, as might be expected from the use of an energy function $\lambda$ times larger than the original. The new dynamics is more costly energetically than the original dynamics, even without accounting for the energy required to inject the added noise \bb{(which scales as $\sigma^2$\c{whitelam2024increasing})}.

However, consistent with our expectation from the previous section, the \bb{reduced work} $\beta_\lambda \av{W_\lambda}_\lambda$ {\em decreases} with increasing  $\lambda$, because the system behaves as if it is subject to a slowed-down version of the original protocol. This decrease is shown in \f{fig2}(b). Indeed, we see that 
\beq
\beta_\lambda \av{W_\lambda}_\lambda = w(\lambda),
\eeq
where the function $w$ is defined in \eqq{ww}, confirming that the $\lambda$-modified system behaves like the original system to which we have applied the protocol $\lambda$ times more slowly, for $\lambda$ times longer. 

Panels (c--e) show that the estimate of $\Delta F$, using \eqq{jarz2}, becomes increasingly precise as $\lambda$ increases. The estimator $J_\lambda$ converges to $\beta \Delta F \approx 0$, and the error measures $\sigma^2_{B}$ and $N_{0.1}$ decrease by orders of magnitude. For fixed trajectory time $\tf=1$, the $\lambda$-modified dynamics (using a simple linear protocol) allows us to estimate $\Delta F$ with much higher precision than does the original dynamics using the optimal protocol.

In \f{fig1}(f) we show the work distributions equivalent to \f{fig1}(b), now for the linear protocol with $\lambda=30$. The work distributions resulting from forward (green) and time-reversed (cyan) protocols cross at $\Delta F \approx 0$. The crossing point is closer to the typical values of the two distributions than in the plots of \f{fig1}(b), and so can be estimated with greater accuracy. The forward distribution weighted by the quantity $\e^{-\beta_\lambda W_\lambda}$ (black dashed) overlaps with the distribution obtained from the time-reversed protocol, for a much larger range than can be determined in \f{fig1}(b).

In \f{fig1}(f), the vertical dotted line denotes the value of $\beta \av{W}_{\rm opt} \approx 8.33$, which is much larger than the mean value $\beta_\lambda \av{W_\lambda}_\lambda \approx 0.8$ obtained using the $\lambda$-modified dynamics. For the latter, for $\lambda=30$, the estimate of $\beta \Delta F$ obtained using the Jarzynski equality is $J_\lambda \approx -0.000931229$, with error parameters $\sigma^2_{B} \approx 0.0394953 $ and $N_{0.1} \approx 400$. Injecting noise has therefore \bb{substantially} increased the precision with which we can estimate $\Delta F$ using the Jarzynski equality.

The microscopic underpinning of this enhanced precision is illustrated in \f{fig3}. There we show time-ordered snapshots of the instantaneous particle-position distribution $\rho(x,t)$, together with the potential $U(x,c(t))$ and the associated Boltzmann distribution 
\beq
\rho_0(x,t) = \frac{\e^{-\beta U(x,c(t))}}{\int {\rm d} x' \, \e^{-\beta U(x',c(t))}}.
\eeq
For the case $\lambda=1$, corresponding to the original dynamics, the protocol results in a far-from-equilibrium \bb{trajectory ensemble}. For the case $\lambda=30$, the system remains close to equilibrium throughout the transformation. In the latter case the dissipated mean \bb{reduced} work $\beta_\lambda \av{W_\lambda}_\lambda -\beta \Delta F$ is smaller, and the resulting free-energy estimate more accurate. The snapshots illustrate the fact that adding noise to the system increases its relaxation rate, and scaling the potential accordingly  leaves the thermodynamics of the system unchanged.

\subsection{Information erasure}

In this section we consider a nonlinear protocol in which a double-well potential is changed into a single-well one, a form of information erasure, with a consequent increase in the free energy of the system. \bb{The work statistics $P(W)$ for this system is not in general Gaussian.} The potential appearing in Eqs.~\eq{jarz} and \eq{jarz2} is now
\bea
\label{pot}
U(x,\cee(t)) &= & \frac{1}{2} \big(x-{\rm sgn}(x-c_0(t))c_1(t)\big)^2 \\ 
 & + & c_0(t) c_1(t)\big({\rm sgn}(x-c_0(t)) +{\rm sgn}(c_0(t))\big), \nonumber
\eea
in units of $\kt$. This potential, borrowed from\cc{Dago-2023-PNAS}, is parameterized by the coefficients $\cee(t)=(c_0(t),c_1(t))$. It has in general a double-well form, shown in \f{fig_erase2}, where $c_0$ tunes the asymmetry and $c_1$ the barrier height. We start with a symmetric double well parameterized by $\cee=(0,5)$, and impose the basic protocol of\cc{Dago-2023-PNAS} (we omit the final-time restoration of the double-well form of the potential). This protocol squeezes the double wells together and translates the resulting single well to the left, as shown in \f{fig_erase2}. In \f{fig_erase1}(a) we show the protocol $\cee(t)$.

In \f{fig_erase1}(b) we show the mean fraction of particles that at time $\tf=1$ are found to the left of the origin. Averages are computed over $10^5$ independent trajectories of the $\lambda$-modified dynamics \eq{lang2}, with $\lambda=1$ corresponding to the case of the original dynamics. Particles begin in thermal equilibrium with respect to the double-well form of the potential, and so with equal likelihood begin either side of the origin (see \f{fig_erase2}). The imposed protocol is designed to bring the particle to the left of the origin, regardless of its starting state, so erasing the information contained in the starting state. However, for the chosen trajectory length $\tf=1$, the particle cannot respond rapidly enough to the driving, and erasure is only about 87\% successful. 

One way to make erasure more effective is to impose a more efficient protocol $\cee(t)$\c{proesmans2020optimal,zulkowski2014optimal,barros2024learning}; another is to increase $\lambda$. As shown in the figure, increasing $\lambda$ increases the erasure probability, which reaches unity for $\lambda \gtrsim 3$. The system behaves as if the erasure protocol were imposed more slowly, allowing the particle to more closely follow the potential as it changes. This effect can be seen in \f{fig_erase2}.

As before, increasing $\lambda$ also allows us to estimate $\beta \Delta F$ accurately, even for the short trajectory length $\tf=1$. The energy barrier separating the initial double wells is $c_1(0)^2/2=12.5$, in units of $\kt$, and so the free-energy difference between initial and final forms of the potential is, to a very good approximation, $\beta \Delta F = \ln2$. This result is closely related to the Landauer observation that erasing one bit of information at temperature $T$ costs at least $\kt \ln 2$ units of work\c{landauer1961irreversibility,bennett1985fundamental}. We show, in panels (c) and (d) of \f{fig_erase1}, that as $\lambda$ increases the free-energy estimator $J_\lambda$ tends to $\ln 2$ and the block-average variance $\sigma_B^2$ diminishes. The faster relaxation engineered by the added noise allows the particle to remain close to equilibrium throughout the transformation, reducing the dissipated \bb{reduced} work and the fluctuations associated with the Jarzynksi free-energy estimator.

\section{Conclusions}
\label{conclusions}

Estimating free-energy differences from nonequilibrium work measurements is often hindered by large work fluctuations, which impair the convergence of the Jarzynski equality. Existing approaches to improving convergence include reducing the protocol driving rate, or designing protocols that reduce dissipated work or work fluctuations.

Here we have proposed the alternative and counterintuitive strategy of {\em adding} noise to the system in order to enhance the precision of free-energy estimates. By introducing additional stochastic fluctuations and rescaling the system's potential energy \bb{accordingly}, we can increase the system's relaxation rate without changing its thermodynamics, and so reduce the magnitude of work fluctuations and suppress rare-event sampling issues. Using numerical simulations of \bb{two} model systems, we have demonstrated that this approach significantly improves the accuracy of free-energy estimates.

However, the practical applicability of this method is limited, because it depends on our ability to effect a uniform scaling of the potential energy landscape. This is not feasible in many settings, such as experiments involving single biomolecules. We can exert {\em some} control over a biomolecule's potential energy landscape, such as by adding NaCl to a solution containing DNA, but not to the required degree. Systems involving rigid particles in optical traps\c{martinez2012effective,chupeau2018thermal,saha2023information} and thermodynamic computers\c{aifer2024thermodynamic} {\em do} satisfy our requirements: external noise can be added to these systems, and the potential energy scaled in the required way. For instance, the method could be used with thermodynamic computers\c{conte2019thermodynamic,aifer2024thermodynamic} as a way to accurately calculate the integrals in \eq{df}, with $U(\x,\cee(0))$ corresponding to a simple reference state and $\e^{-\beta U(\x,\cee(\tf))}$ the desired integrand. However, for particle-in-trap systems we usually know the free energies of the relevant states, making the method redundant (in this case the method could be used to simply enact a given protocol more quickly\c{whitelam2024increasing}). 

\bb{The same time rescaling could in principle be done for underdamped Langevin dynamics, although the experimental requirements (beyond adding noise) are even more difficult to achieve than in the overdamped case. The underdamped Langevin equation reads \beq
\label{ud1}
m\ddot{x}_i+ \gamma\dot{x}_i =- \frac{\partial U({\bm x})}{\partial x_i}+ \sqrt{2 \gamma \kt} \, \eta_i(t),
\eeq
where first term represents inertial effects ($m$ is mass), the second term is the damping force ($\gamma = 1/\mu$ is the damping coefficient), and the noise term models thermal fluctuations. Rescaling time by a factor $\lambda \geq 1$ in \eq{ud1} results in the equation
\beq
\label{ud2}
m\ddot{x}_i+ \lambda \gamma\dot{x}_i =- \lambda^2 \frac{\partial U({\bm x})}{\partial x_i}+ \sqrt{2 \lambda^3 \gamma \kt} \, \eta_i(t).
\eeq
We can produce \eq{ud2} from \eq{ud1} -- and so effectively increase the time constant of the system -- by adding to \eq{ud1} a Gaussian white noise with zero mean and variance $2 (\lambda^3 -1)\gamma \kt$, and rescaling the potential $V$ by a factor of $\lambda^2$. However, we also need to rescale the damping term by a factor of $\lambda$, which in an optical trap setup would require imposing additional frictional forces.}

In general terms, our results indicate that noise engineering could be a useful tool for enhancing thermodynamic measurements and computations, and suggest that the notion of optimal control could be broadened to include control of noise, which is now experimentally feasible, as well as control of the potential. 

\section{Acknowledgments} I thank Corneel Casert for comments on the paper. This work was performed at the Molecular Foundry at Lawrence Berkeley National Laboratory, supported by the Office of Basic Energy Sciences of the U.S. Department of Energy under Contract No. DE-AC02--05CH11231.

\appendix

\section{Insight into the convergence of \eqq{jarz} assuming Gaussian work statistics}
\label{gauss}

The convergence properties of the Jarzynski equality, \eqq{jarz}, can be \bb{illustrated} by assuming that the work distribution $P_{\rm F}(W)$ is Gaussian with mean $\av{W}$ and variance $\sigma^2$. In this case the expectation in \eq{jarz} can be written 
 \bea
 \label{expect}
 \av{\e^{-\beta W}} &=& \e^{-\beta \av{W} + \frac{1}{2} \sigma^2 \beta^2} \\
 &\times& \frac{1}{\sqrt{2 \pi \sigma^2}} \int_{-\infty}^\infty {\rm d} W \, \e^{-\frac{1}{2 \sigma^2} (W-\av{W} + \sigma^2 \beta)^2} \nonumber.
 \eea
 Two things are apparent from \eqq{expect}. First, in order to evaluate the integral by trajectory sampling, we must collect good statistics at the atypical work value $W_0 = \av{W} -\sigma^2 \beta$ that dominates the integral. This value occurs with probability $P_{\rm F}(W_0) \sim \e^{-\sigma^2 \beta^2/2}$, meaning that for good sampling of \eq{jarz} we must generate a number of trajectories $N_{\rm traj} \sim P_{\rm F}(W_0)^{-1}$  that increases exponentially with the variance $\sigma^2$ of \bb{work fluctuations}. Second, by equating the right-hand side of \eq{jarz} with the right-hand side of \eq{expect} (noting that the second line is unity), we see that
 \beq
 \label{link}
\beta  \sigma^2=2 (\av{W}-\Delta F),
 \eeq
 which means that the work distribution $P_{\rm F}(W)$ with the smallest variance corresponds to the distribution with the smallest mean dissipated work $\av{W}-\Delta F$\c{geiger2010optimum}.
 
For the trap-translation problem considered in the main text, the minimum-work protocol has Gaussian work statistics, and is the protocol that ensures best convergence of \eq{jarz}\c{geiger2010optimum}. In general, work distributions are not exactly Gaussian, and so the protocol that ensures best convergence of \eq{jarz} is not necessarily the protocol with the smallest mean dissipated work.


%

\end{document}